\documentclass[aps,prl,showpacs,twocolumn,groupedaddress,amssymb]{revtex4}

\usepackage{amsmath}
\usepackage{graphicx}
\usepackage{dcolumn}
\usepackage{bm}
\usepackage{subfigure}
\usepackage{amssymb}
\usepackage{dcolumn}

\begin{document}

\title{Ignition and Front Propagation in
Polymer Electrolyte Membrane Fuel Cells}

\author{J. B. Benziger}
\author{E.-S. Chia}
\author{Y. De Decker}
\author{I. G. Kevrekidis}

\affiliation{Department of Chemical Engineering, Princeton
University, Princeton, NJ 08544, USA}

\date{\today}

\begin{abstract}
Water produced in a Polymer Electrolyte Membrane (PEM) fuel cell
enhances membrane proton conductivity; this positive feedback loop
can lead to current ignition.
Using a segmented anode fuel cell we study the effect of gas phase
convection and membrane diffusion of water on the spatiotemporal nonlinear dynamics
--localized ignition and front propagation-- in the cell.
Co-current gas flow causes ignition at the cell outlet, and membrane diffusion
causes the front to
slowly propagate to the inlet; counter-current flow causes ignition
in the interior of the cell, with the fronts subsequently spreading
towards both inlets.
These instabilities critically affect fuel cell performance.
\end{abstract}


\maketitle

Fuel cells constitute a reliable and environmentally friendly
alternative energy source.
Beyond steady state operation, understanding of their startup and transient dynamic
behavior is crucial to their use in {\it variable load} automotive applications
\cite{Martin05,Fronk05}.
These transient dynamics are intensely nonlinear due to a positive
feedback loop: we recently demonstrated that water generated in a polymer electrolyte membrane
(PEM) fuel cell {\it increases} proton transport exponentially, which
``ignites" the current \cite{moxley03,chia04}.
There is a strong analogy to the positive feedback loop leading to
nonlinear dynamics such as ignition and spatiotemporal front
propagation in exothermic chemical reactions.
There, a product of the reaction (heat) enhances the reaction
rate by raising the temperature \cite{hotspots}.
In autohumidified PEM fuel cells, the positive feedback between membrane water
activity and proton conductivity  is known to cause steady state
multiplicity \cite{benziger04}.
%
\begin{figure}[t]
\begin{flushright}
\includegraphics[angle=-90,width=.99\columnwidth]{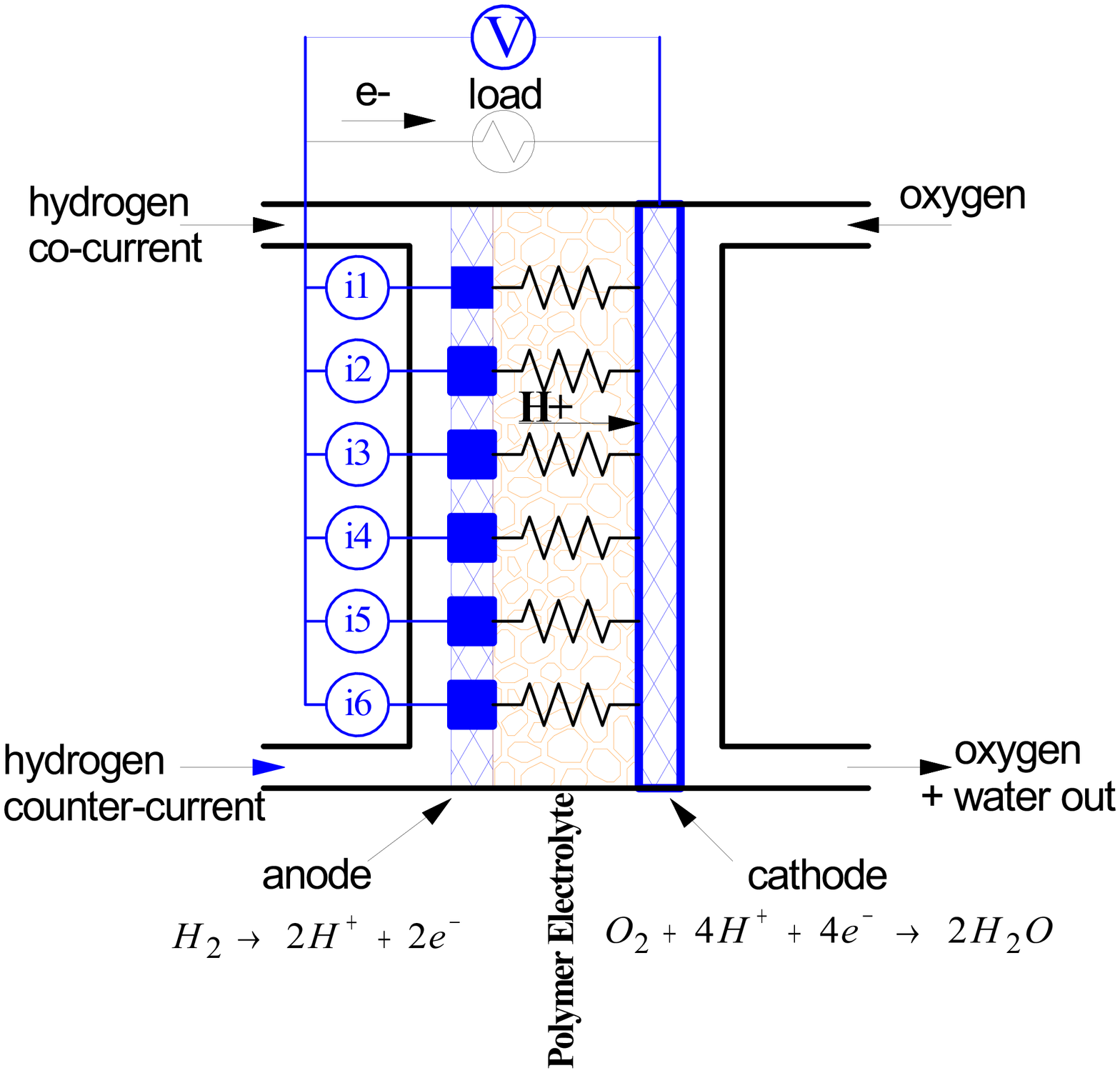} \vspace{0cm}
\caption{ \label{fig1}
(Color online) Schematic of a Hydrogen-Oxygen PEM fuel cell
including internal and external equivalent circuit elements.
Hydrogen molecules dissociatively adsorb at the anode and are
oxidized to protons.
Electrons travel through an external load resistance to the cathode,
while protons diffuse through the PEM under an electrochemical
gradient to the cathode.
Oxygen molecules adsorb at the cathode, are reduced, and react with
the protons to produce water.
The product water is absorbed into the PEM, or evaporates into the
gas streams at the anode and cathode.}
\end{flushright}
\end{figure}

Here we present {\em spatiotemporal} nonlinear dynamic
phenomena (current ignition and current
density front propagation along flow channels) in a two-dimensional
fuel cell.
PEM fuel cells typically have flow channels that distribute the fuel
(hydrogen) across the anode and oxidizer (oxygen) across the
cathode (Fig. \ref{fig1}) \cite{Promislow04}.
Longitudinal water gradients in the membrane produce a
sharp current density front that propagates in time along the
channel \cite{VanZee0003}.
To design and control variable load, {\it dynamic} fuel cell operation,
the mechanism underlying such dynamics must be understood.

\begin{figure}[t]
\begin{center}
\includegraphics[width=.8\columnwidth]{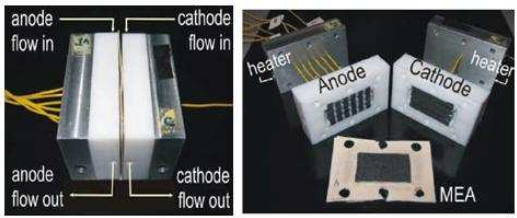}
\caption{ \label{fig2} (Color online) The segmented anode fuel cell.
The anode electrode was broken into six individual elements
separated by Teflon spacers.
Current through each element was measured independently.
A membrane-electrode-assembly employing 2
ETEK electrodes with carbon supported Pt catalyst and a Nafion 115
membrane was placed between anode and cathode. }
\end{center}
\end{figure}

{\bf Experimental.} Fig. \ref{fig2} shows a photograph of a simplified fuel
with a segmented anode, that permits current profile measurements.
The current through each segment of the anode and the voltage drop
across the external load resistor are recorded as a function of
time.
The cell operates at atmospheric pressure;
the flow rates of hydrogen at the anode and oxygen at the
cathode are maintained with mass flow controllers.
%
%

The fuel cell is placed between two temperature controlled aluminum
blocks.
%
There are six segments to the fuel cell, each with an active area of
$0.5$ cm$^2$.
It is preconditioned with
a 20 $\Omega$  load at 60 $^0$C and
flow rates of 3.5 sccm H$_2$ at the anode and O$_2$ at the cathode
for 8-12 hours.
The current in each segment was $<$ 1 mA after
preconditioning (extinction).

\begin{figure}[t]
\begin{center}
\includegraphics[width=1.2\columnwidth]{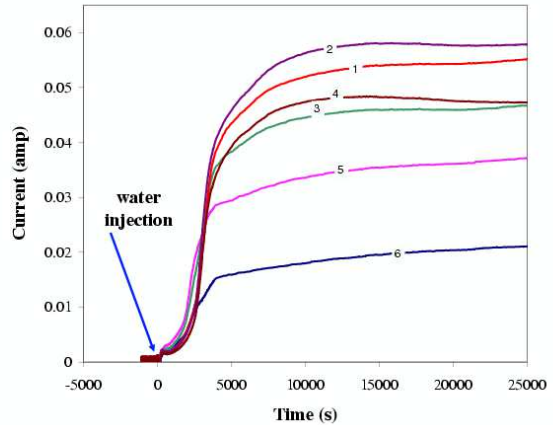}
\caption{ \label{fig3} Fuel cell ignition by water injection.
The fuel cell was preconditioned so as to dry the membrane; the current in
each segment of the anode was then $<$ 1 mA.
At t = 0, 100  $\mu$L of water
were injected into the anode (hydrogen) feed.
After the water was
absorbed into the membrane, the fuel cell currents rose dramatically
(ignited), and remained so.
The numbers refer to the currents
in the segments in order - 1 is at the anode inlet and 6 is at the
anode outlet. }
\end{center}
\end{figure}

The external load was then reduced to 0.25 $\Omega$
keeping temperature and flow rates fixed.
The current in each
segment remained at $<$ 1 mA for more than 4 hours,
at which time 100 $\mu$L of water
was injected into the anode feed stream.
The current response for each segment after the water
injection is shown in Fig. \ref{fig3}.
The injection of the water ``ignited" the fuel cell current.
Before water injection the resistance of the dry membrane for proton
conduction was very high ($>$ 10 k$\Omega$ cm$^{-2}$), limiting
the current and hence the water production.
The injected water was absorbed into
the membrane, decreasing the membrane resistance to
$\sim$ 1 $\Omega$ cm$^{-2}$; this increased the currents to $\sim$
100 mA and made the fuel cell self-sustaining.

To explore the
location and dynamics of ignition in the flow channels, the fuel cell
was extinguished as described above.
The temperature was then reduced to 25 $^0$C, decreasing the water
removal, and the load resistance reduced to 0.25 $\Omega$,
increasing the water production.
Current and voltage were
recorded every 100 s for a period of 20 hours after these changes.
For the measurements reported here, the fuel cell was
placed with the flow channels running vertically.
Co-current
measurements were made with both H$_2$ and O$_2$ flows going from
top to bottom; for counter-current measurements, the oxygen flow
went from top to bottom and the hydrogen flow from bottom to top.
%

\begin{figure}[t]
\begin{center}
\includegraphics[width=.99\columnwidth]{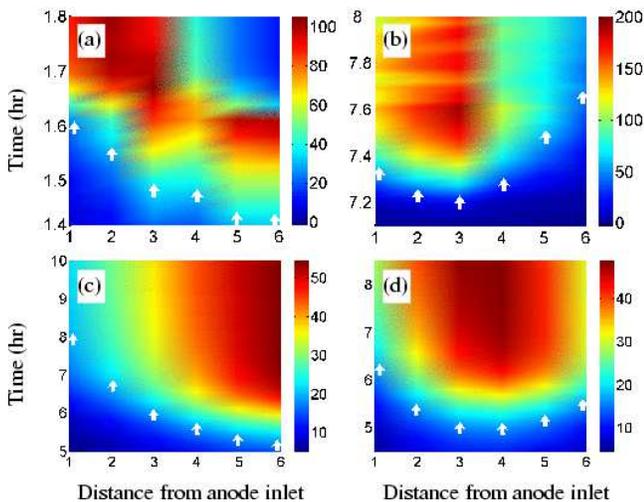}
\caption{ \label{fig4} (Color online) A comparison of the
experimental and computed currents for co-current and
counter-current flow of hydrogen and oxygen in a segmented anode PEM
fuel cell.
The color scale is for current through each anode segment in mA. (a)
experimental co-current (b) experimental counter-current (c)
computed co-current (d) computed counter-current.
For the simulations,
the flow rate for H$_2$ and O$_2$ is 3.5 and 6.5 mL min$^{-1}$ respectively
and $k_m=3\times 10^{-6}$ mol s$^{-1}$.}
\end{center}
\end{figure}

The ignition and subsequent current distributions along the flow
channels as functions of time are shown in Figs. \ref{fig4}(a) and
\ref{fig4}(b).
An induction period of hours was required before any significant
current was measured; the length of this induction period depended
on how dry the fuel cell membrane was before startup (i.e. before
reducing the load resistance.)
For co-current flow (Fig.\ref{fig4}(a))
ignition first occurred at the outlet of the fuel cell; the current in
anode element 6, near the outlet, rose from $<$ 1 mA to $\sim$ 100
mA over a period of 5 min.
A current ignition front then propagated
from element 6 to element 1 over a period of 15-20 minutes; as the
current front propagated towards the entrance of the flow channels,
the current at the exits of the flow channels dropped.
With counter-current flow ignition first occurred at element 3, at the
interior of the flow channel (Fig. \ref{fig4}(b)).
From the center, the
ignition fronts ``fanned" outwards, but the highest current always
occurred in the center of the flow channel.
The induction period
was also longer with counter-current flow than with co-current flow.
The long-term, total steady state steady state current was roughly
double for counter-current (as opposed to co-current) flow.

{\bf Modeling.} We have been able to capture the basic physics of
ignition dynamics in a simple lumped {\it differential}, single anode fuel
cell \cite{benziger04}; to capture {\em localized} ignition and spatiotemporal front
propagation the model is extended to a series of differential elements.
The key features of
the model are the water inventory in the polymer electrolyte, the
transverse proton conductivity from the anode to the cathode and the
longitudinal water transport through the membrane.
Water sorbed into the
polymer ionizes sulfonic acid groups, facilitating proton transport;
water sorption by the PEM is limited by the total number
of sulfonic acid groups in the membrane.
Water is sorbed into, or
desorbed from, the membrane depending on the balance between water
evaporating into the gas flow channels and water produced by the
fuel cell.
The multilayered membrane-electrode-assembly in PEM fuel
cells results in complex intra- and interlayer transport processes.
Since the flow
channels are much longer than the membrane thickness (5 cm vs.
0.0125 cm), we make the simplifying assumption
that water activity in the membrane is
in local equilibrium with water activity in the gas flow
channel above it, and the only gradients are longitudinal {\it along} the channel.

The water balance in each differential element $j$ of the membrane
is given by eq. \ref{eq1} for co-current gas flow ($j=1$ to 6; $j=0$
and $j=7$ denote the anode inlet and outlet, respectively; the
counter-current case easily follows); the inventory is balanced by
water produced (1/2 the proton current), water convected in the gas
flow, and longitudinal water diffusion (described by a lumped mass
transfer coefficient between differential elements).
Eq. \ref{eq2} is an empirical fit to the number
of water molecules associated with each sulfonic acid group, $\lambda$, as a
function of water activity $\alpha_w$ in a Nafion 115 membrane \cite{yang04}.
\begin{eqnarray} \nonumber
{}&{}& \left[N_{{\rm SO}_3}\, \frac{d\lambda(j)}{da_w(j)} + \left(
V_A+V_C\right)\, \frac{P_w^0}{RT}\right] \, {\dot{a}_w(j)} = \\
\nonumber {}&{}& \frac{i(j)}{2\mathcal{F}} + \left[
F_A(j-1)+F_C(j-1) \right] P_w(j-1) \\ \nonumber {}&{}& - \left[
F_A(j)+F_C(j) \right]P_w(j) +k_m\left[a(j+1)+a(j-1) \right. \\
\label{eq1} {}&{}&  \left. -2a(j) \right]
\end{eqnarray}
\begin{eqnarray} \nonumber
\lambda(j) &=& 14.9\, a_w(j) - 44.7\, a_w^2(j)+ 70 \, a_w^3(j)
\\ \label{eq2} &{}&-26.5 \, a_w^4(j) - 0.446\, a_w^5(j)
\end{eqnarray}
%
We assume that the total gas pressure $P_T$ is fixed, and the local water
activity in the membrane is in equilibrium with the local water
partial pressure $P_w$, i.e. $P_w(j)=a_w(j)\, P_w^0$ and $P_{\rm H_2}(j)=P_{\rm
O_2}(j)=P_{T}-P_w(j)$; $P_w^0$ is the water vapor pressure at this temperature.
The molar flow rates change along the flow
channel as water is formed; the molar flows are given by
$F_A(j)=F_A(j-1)-i(j)/4\mathcal{F}$ and $F_C(j)=F_C(j-1)$; the
subscript $A$($C$) corresponds to the anode(cathode), $i(j)$ is
the local current and $\mathcal{F}$ the Faraday constant.
Lastly,
we assume that the local potential $V_{FC}$ between the anode and cathode is dictated
by the thermodynamic driving force as in eq. \ref{eq3}.
This neglects
interfacial potential drops, which results in the predicted currents
being about 20$\%$ larger than found in real fuel cells.
%
\begin{eqnarray} \label{eq3}
V_{FC}(j) = 1.23  - \frac{RT}{4\mathcal{F}}\ln\left[\frac{P_{{\rm
H}_2}^2(j)\, P_{{\rm O}_2}(j)}{P^3_{\rm atm}\, a_w^2(j)}
\right]\qquad \left[ V\right]
\end{eqnarray}
%
Based on the equivalent electrical circuit, the differential
elements are electrically connected {\it in parallel} to each other.
The
voltage across the external load resistance thus depends on the
total current produced by all elements; the local current is given
by eq. \ref{eq4}.
The local membrane resistance, $R_M(j)$, depends of
the local water content in the membrane.
For a Nafion 115 membrane
employed in this fuel cell the membrane resistance as a function of
water activity is given by eq. \ref{eq5} \cite{yang04}.
%
\begin{eqnarray} \label{eq4}
i(j) = \frac{V_{FC}(j)-R_L\, \sum_{k\neq j}i(k)}{R_M(j)+R_L}\qquad
\left[ A \right]
\end{eqnarray}
\begin{eqnarray} \label{eq5}
R_M(j) =5 \times10^{5}\exp{\left(- 14\, a_w^{0.2}\right)}
\qquad \left[ \Omega \right]
\end{eqnarray}
%
In a {\it single}, differential PEM fuel cell ignition occurs when the
initial water content in the membrane is sufficient for water
production to exceed water removal by convection.
Here ignition will occur when (and where!) the water production exceeds water removal,
i.e. when (and where) the right hand side of eq. \ref{eq1} becomes greater
than zero.
Water production depends on the load resistance and the
membrane resistance.
A dry membrane has a resistance of 500 k$\Omega$ cm$^{-2}$, limiting
the current density to a maximum of 2.4 $\mu$A cm$^{-2}$.
According to eq. \ref{eq1} the feed flow rates would
have to be $<$ 0.1 mL hr$^{-1}$ at a fuel cell temperature of 60
$^{0}$C for water production to be greater than water removal and
ignite the fuel cell.
Absorption of 10  $\mu$L cm$^{-2}$ of water
into the membrane reduces its resistance to $\sim$ 10 $\Omega$
cm$^{-2}$ and the maximum current density is 100 mA cm$^{-2}$,
sufficient for water production to exceed water
removal, and the fuel cell to ignite as shown in Fig. \ref{fig3}.

The key elements that account for ignition are (1) an exponential
dependence of proton conductivity in the PEM with membrane water
content and (2) the dynamics of water uptake into the PEM.
The
location of ignition and front propagation are consequences of (1)
convection of water produced towards the cell downstream, where it can accumulate and
(2) diffusion of water upstream through the polymer membrane.
Figs. \ref{fig4}(c) and (d) show the {\it simulated} current
profiles for co-current and counter-current flow with $T=47\, ^0{\rm
C}$, $R_L=5\, \Omega$.
The model captures both the
ignition and the front propagation, though it overpredicts the
experimentally observed currents.

The model is only semi-quantitative because it neglects finite water mass
transfer rates into the membrane and from the membrane into
the gas phases.
It also neglects the
effects of condensing liquid water, hindering gas transport from the flow
channels to the membrane/electrode interface.
More detailed models
that incorporate these effects can give quantitative fits to
experimental results; yet the added complexity does not
significantly enhance physical understanding.

For co-current flow the water produced upstream is conducted towards
the outlet, where it slowly accumulates in the membrane; when the
water content increases to the point where the local membrane
resistance becomes comparable to the external load resistance the
current starts increasing rapidly, hydrating the membrane and causing
ignition at the outlet of the flow channel.
Upon ignition the water
activity in the membrane approaches unity.
Water is then transported
upstream through diffusion in the membrane itself, causing upstream
propagation of the ignition.
The model did not capture the observed eventual decrease
in the downstream current, after the ignition propagated to the inlet
of the fuel cell.
This decrease is due to {\em condensing water}
accumulating in the cathode, inhibiting oxygen transfer to the
catalyst and reducing the current.

In counter-current flow, water formed at the cathode is convected
towards the anode inlet in the cathode flow channel.
It is also transported {\it across the membrane} to the
anode, where it is convected towards the anode outlet.
Water
accumulates fastest towards the middle of the flow channels,
resulting in an interior ignition point.
The ratio of the flow rates
between the anode and cathode affects ignition point location; a
relative increase of the anode flow rate shifts ignition towards the
anode outlet.
After ignition, water starts to accumulate locally in the membrane.
The
transport of water {\it through} the membrane from high concentration at
the middle of the flow channel towards the outlets results in the
``fanning out" of the ignition fronts.

Liquid water was observed leaving the flow channels 30-40 minutes
after ignition.
Ignition takes the fuel cell from very low water
activity to water activity of unity: condensation of liquid water in the flow
channels.
Gravity plays a key role in how such liquid water moves
through the flow channels; the operation changes dramatically if gas
flow in the channel is counter to gravity driven liquid water flow.
When the fuel cell was vertical, gravity caused the liquid to drain,
and permitted good access for the reactants from the flow channels
to the electrode/electrolyte interface.
When the fuel cell was horizontal the initial ignition phenomena
were similar to those reported in Fig. \ref{fig3}; after ignition,
however, large fluctuations in the local current density appear to
correlate with water droplets exiting the cell.
In the horizontal orientation, liquid water condensing in the
flow channels could partially block flow.
The liquid
drops were pushed along the flow channels by the flowing gas, but in
an irregular fashion, that gave rise to large fluctuations in the
local current density.
Transport of liquid water in the flow
channels is not accounted for at our level of modeling.
%
%
The model can thus capture ignition
and front propagation, but will break down at longer times, when
liquid water floods the cathode gas diffusion layer (GDL)
and starts entering the flow channels.

We have demonstrated how the water exponentially increases proton
conductivity in polymer electrolyte membranes
leading to ignition of the current in PEM fuel cells.
Water can be
injected to a fuel cell to ignite the current, just like a match can
be struck to ignite a flame.
Transport of water laterally, coupled
with the exponential increase in proton conductivity produces
current fronts that propagate along flow channels, just like flame
fronts.
The positive feedback loop between water production and
increased proton conductivity of the electrolyte membrane in PEM
fuel cells is analogous to exothermic chemical reaction ignition and
flame propagation, yet ``water fans the flame" in
PEM fuel cells!
Front propagation depends on flow
configurations creating different front ignition and propagation
patterns.
Understanding the parametric dependence and time scales of
these phenomena is a vital component of fuel cell design,
non-steady state operation and control.

{\bf Acknowledgements} We thank the National Science Foundation (CTS
-0354279 and DMR-0213707) for support of this work. E. Chia thanks
the Princeton University Program in Plasma Science and Technology,
and U.S. Department of Energy Contract No. DE-AC02-76-CHO- 3073 for
fellowship support. Y. De Decker thanks the Belgian American
Educational Foundation for financial support.

\end{document}